\documentclass[3p,times]{elsarticle}

\usepackage{ecrc}

\usepackage{epstopdf}

\volume{00}

\firstpage{1}

\journalname{Nuclear Physics A}

\runauth{Author1 et al.}

\jid{nupha}

\jnltitlelogo{Nuclear Physics A}

\usepackage{graphicx}
\usepackage{amsmath,amssymb}
\usepackage{epsfig}
\usepackage{subfigure}

\begin{document}

\begin{frontmatter}



\title{Heavy Flavor Dynamics in QGP and Hadron Gas}

\author[duke]{Shanshan Cao}
\address[duke]{Department of Physics, Duke University, Durham, NC 27708, USA}
\author[ccnu]{Guang-You Qin}
\address[ccnu]{Institute of Particle Physics and Key Laboratory of Quark and Lepton Physics (MOE), Central China Normal University, Wuhan, 430079, China}
\author[duke]{Steffen A Bass}



\begin{abstract}
We study heavy flavor evolution in the quark-gluon plasma matter and the subsequent hadron gas created in ultrarelativistic heavy-ion collisions. The motion of heavy quarks inside the QGP is described using our modified Langevin framework that incorporates both collisional and radiative energy loss mechanisms; and the scatterings between heavy mesons and the hadron gas are simulated with the UrQMD model. We find that the hadronic interaction further suppresses the $D$ meson $R_\mathrm{AA}$ at high $p_\mathrm{T}$ and enhances its $v_2$. And our calculations provide good descriptions of experimental data from both RHIC and LHC. In addition, we explore the heavy-flavor-tagged angular correlation functions and find them to be a potential candidate for distinguishing different energy loss mechanisms of heavy quarks inside the QGP.

\end{abstract}


\end{frontmatter}



\section{Introduction}
\label{sec:introduction}

Heavy flavor hadrons serve as excellent probes of the quark-gluon plasma (QGP) matter created in ultrarelativistic heavy-ion collisions. Experimental observations at both RHIC and LHC have revealed many interesting data on heavy flavor mesons and their decay electrons especially their small values of $R_\mathrm{AA}$ and large values of $v_2$ which are comparable to those of light hadrons \cite{Adamczyk:2014uip,Adare:2010de,ALICE:2012ab,Abelev:2013lca}. This seems inconsistent with earlier expectation of the mass hierarchy of parton energy loss and thus requires a thorough understanding of heavy flavor dynamics.

Various theoretical frameworks have been developed to study the heavy flavor such as the Boltzmann-based parton cascade model \cite{Uphoff:2012gb}, the linearized Boltzmann transport \cite{Gossiaux:2010yx}, and the Langevin approach \cite{Cao:2013ita}. In this work, we construct a framework that describes the full-time evolution of heavy flavor in heavy-ion collisions. The motion of heavy quarks inside the QGP is described with our improved Langevin equation that incorporates both collisional and radiative energy loss \cite{Cao:2012au,Cao:2013ita}, the hadronization process is calculated with our hybrid model of fragmentation plus coalescence \cite{Cao:2013ita}, and the hadronic interaction is simulated with the UrQMD model \cite{Bass:1998ca}. With this  framework, we not only provide good descriptions of single heavy flavor spectra, but also explore the angular correlation functions of heavy flavor pairs and find them sensitive to different energy loss mechanisms of heavy quarks inside the QGP.

\section{Evolution of Heavy Flavor in Heavy-ion Collisions}
\label{sec:evolution}

Since most heavy quarks are produced in the primordial stage of collisions, we use a Monte-Carlo Glauber model to initialize their positions and pQCD calculations for their momenta. The CTEQ parametrization is adopted for the parton distribution functions which are further modified by the EPS09 parametrization to include the nuclear shadowing effect. The bulk matter (QGP) -- is initialized with either the Glauber or the KLN model.

\begin{figure}[tb]
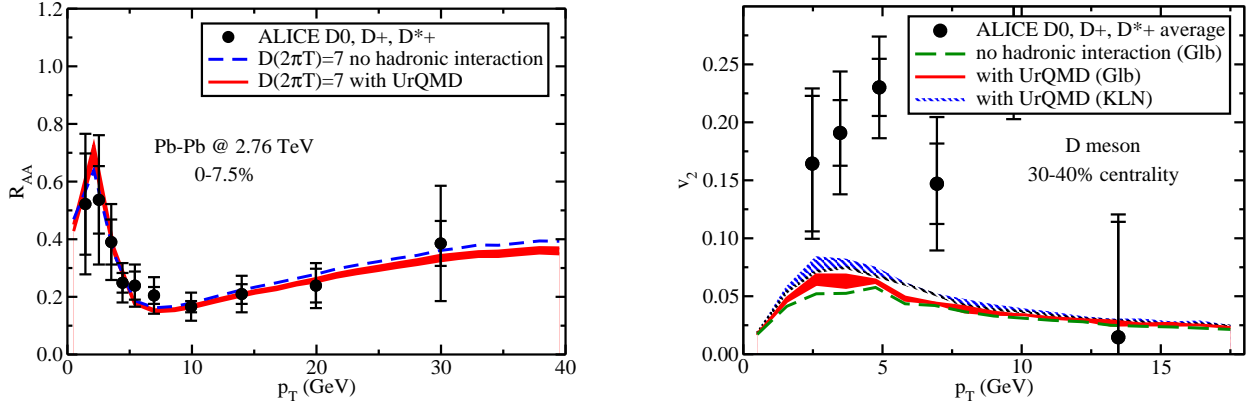

 \subfigure{\epsfig{file=UrQMD-RAA_LHC-0-7d5.eps, width=0.455\textwidth, clip=}}
 \hspace{0.07\textwidth}
 \subfigure{\epsfig{file=UrQMD-v2_LHC-30-50.eps, width=0.455\textwidth, clip=}}
 \caption{(Color online) $D$ meson suppression (left) and flow (right) in the LHC experiment.}
 \label{fig:D-LHC}
\end{figure}

In the QGP stage, the space-time evolution of the bulk matter is simulated by a (2+1)-D viscous hydrodynamic model \cite{Qiu:2011hf} and the in-medium evolution of heavy quarks is described using our improved Langevin equation \cite{Cao:2013ita}:
\begin{equation}
\frac{d\vec{p}}{dt}=-\eta_D(p)\vec{p}+\vec{\xi}+\vec{f_g}.
\end{equation}
The first two terms on the right are the drag and the thermal forces describing the multiple scatterings of heavy quarks off light partons in the medium background, and are related by the fluctuation-dissipation theorem $\eta_D(p)=\kappa/(2TE)$. Here the momentum space diffusion coefficient $\kappa$ is defined in $\langle\xi^i(t)\xi^j(t')\rangle=\kappa\delta^{ij}\delta(t-t')$. The third term $\vec{f_g}=-d\vec{p_g}/dt$ is introduced as the recoil force exerted on the heavy quark while it radiates gluons. The probability of gluon radiation and the energy-momentum distribution of the emitted gluons are calculated according to the following gluon distribution function taken from the Higher-Twist energy loss formalism \cite{Zhang:2003wk}:
\begin{equation}
\label{Wang}
\frac{dN_g}{dx dk_\perp^2 dt}=\frac{2\alpha_s(k_\perp)}{\pi} P(x) \frac{\hat{q}}{k_\perp^4} \textnormal{sin}^2\left(\frac{t-t_i}{2\tau_f}\right)\left(\frac{k_\perp^2}{k_\perp^2+x^2 M^2}\right)^4,
\end{equation}
where $\hat{q}$ is the gluon transport coefficient, $x$ is the fractional energy carried by radiated gluon, and $k_\perp$ is its transverse momentum. We relate the spatial diffusion coefficient of heavy quark $D$ with $\kappa$ and $\hat{q}$ via $D=2T^2/\kappa$ and $\hat{q}=2\kappa C_A/C_F$, so that in end only one free parameter exists. It has been shown in \cite{Cao:2013ita} that while the collisional energy loss dominates the low $p_\mathrm{T}$ region of heavy quark, gluon radiation dominates at high $p_\mathrm{T}$.

After the heavy quarks traverse the QGP medium, they hadronize based on our hybrid model of fragmentation plus coalescence. And it has been shown in \cite{Cao:2013ita} that while the fragmentation mechanism dominates heavy meson formation at high $p_\mathrm{T}$, the heavy-light quark coalescence significantly enhances its production at medium $p_\mathrm{T}$. The decay of the bulk matter is calculated according to the Cooper-Frye formula.  

Finally, both heavy mesons formed from heavy quarks and soft hadrons decayed from the QGP are fed into the UrQMD model for the hadronic interactions. One of the most important ingredients of the UrQMD model are the hadronic scattering cross sections. To simulate the interactions of $D$ mesons with the hadron gas, we introduce the scattering cross sections of charm mesons with pions and rho mesons calculated in Ref. \cite{Lin:2000jp} into UrQMD.

\begin{figure}[tb]
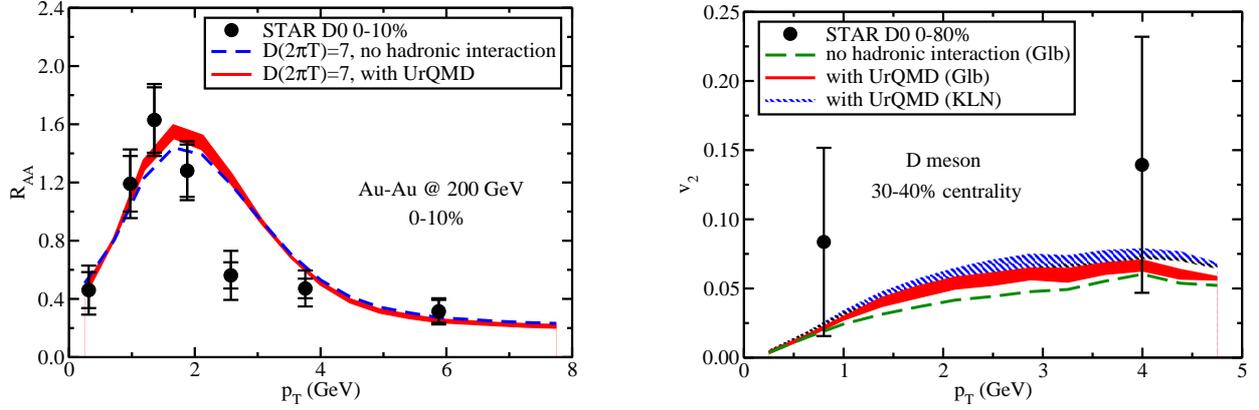

 \subfigure{\epsfig{file=UrQMD-RAA_RHIC-0-10.eps, width=0.455\textwidth, clip=}}
 \hspace{0.07\textwidth}
 \subfigure{\epsfig{file=UrQMD-v2_RHIC-30-40.eps, width=0.455\textwidth, clip=}}
 \caption{(Color online) $D$ meson suppression (left) and flow (right) in the RHIC experiment.}
 \label{fig:D-RHIC}
\end{figure}

We show our calculations of the $D$ meson $R_\mathrm{AA}$ in the left panel of Fig.\ref{fig:D-LHC} for the LHC experiments. We observe that due to additional energy loss that the $D$ meson suffers inside the hadron gas, its $R_\mathrm{AA}$ is further decreased at large $p_\mathrm{T}$. The error band corresponds to the uncertainty in choosing the cutoff parameter of the hadron form factor for calculating scattering cross sections \cite{Lin:2000jp}. Our calculations provide a good description of the observed $D$ meson $R_\mathrm{AA}$ with an extraction of the gluon transport coefficient $\hat{q}$ of around 2.6~GeV$^2$/fm. In the right panel, we find that additional scatterings of $D$ mesons in an anisotropic hadron gas further enhance the $v_2$ value by over 30\%. Two different initializations of hydrodynamical evolution are also compared which may cause another 30\% difference in the $D$ meson $v_2$. In Fig.\ref{fig:D-RHIC}, we present our calculations for the RHIC experiments, the above conclusions remain and our results are consistent with the experimental data. Centrality and participant number dependences of $D$ meson $R_\mathrm{AA}$ can be found in our presentation as well.

\section{Angular Correlation Functions of Heavy Flavor Pairs}
\label{sec:correlation}

\begin{figure}[tb]
\begin{minipage}{0.47\textwidth}
\includegraphics[width=0.96\textwidth,clip=]{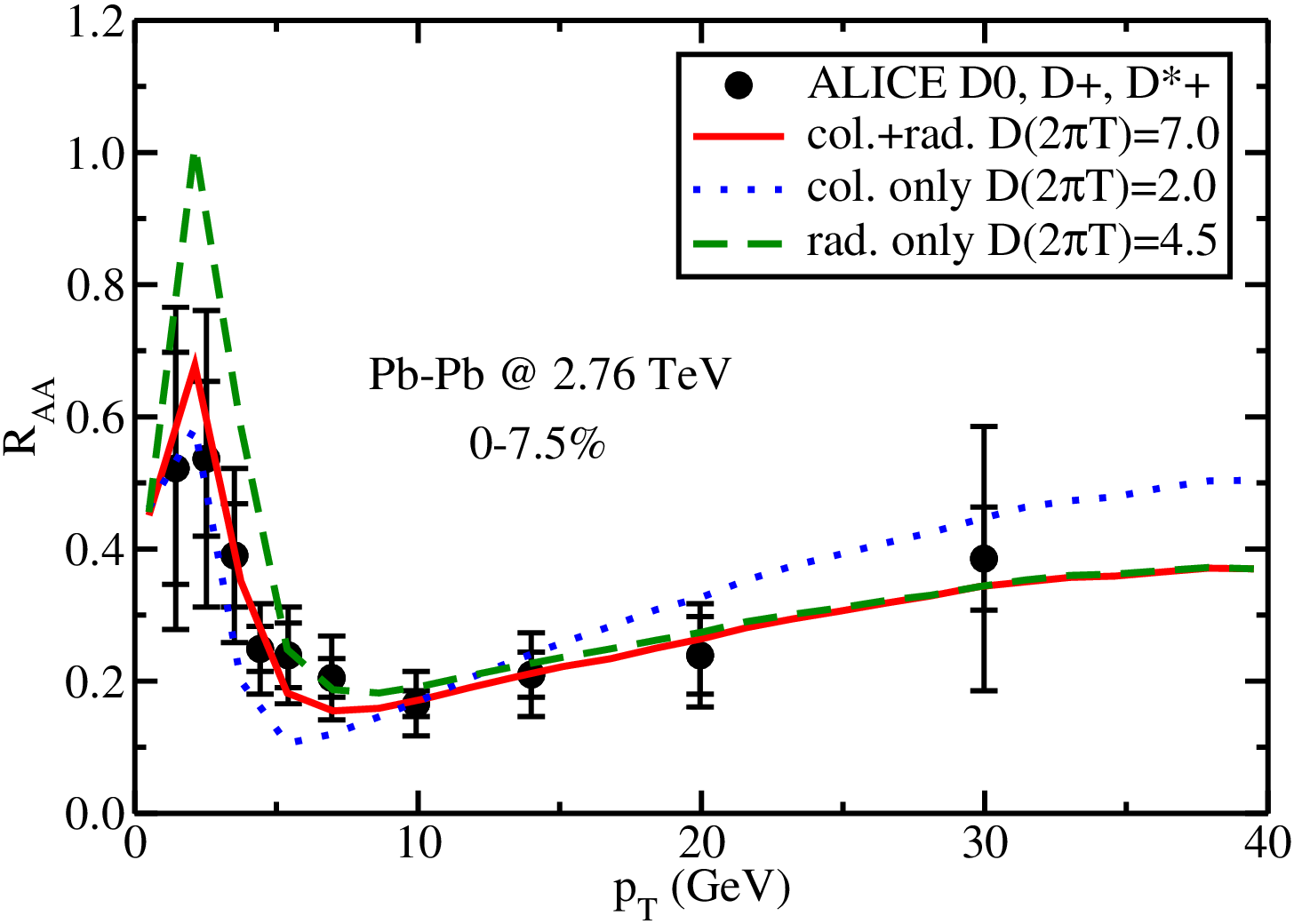}
\caption{\label{fig:RAA_fit_with_each_mechanism}(Color online) Fitting $D$ meson $R_\mathrm{AA}$ with different energy loss mechanisms by tuning the diffusion coefficient.}
\end{minipage}\hspace{0.06\textwidth}
\begin{minipage}{0.47\textwidth}
\includegraphics[width=0.96\textwidth,clip=]{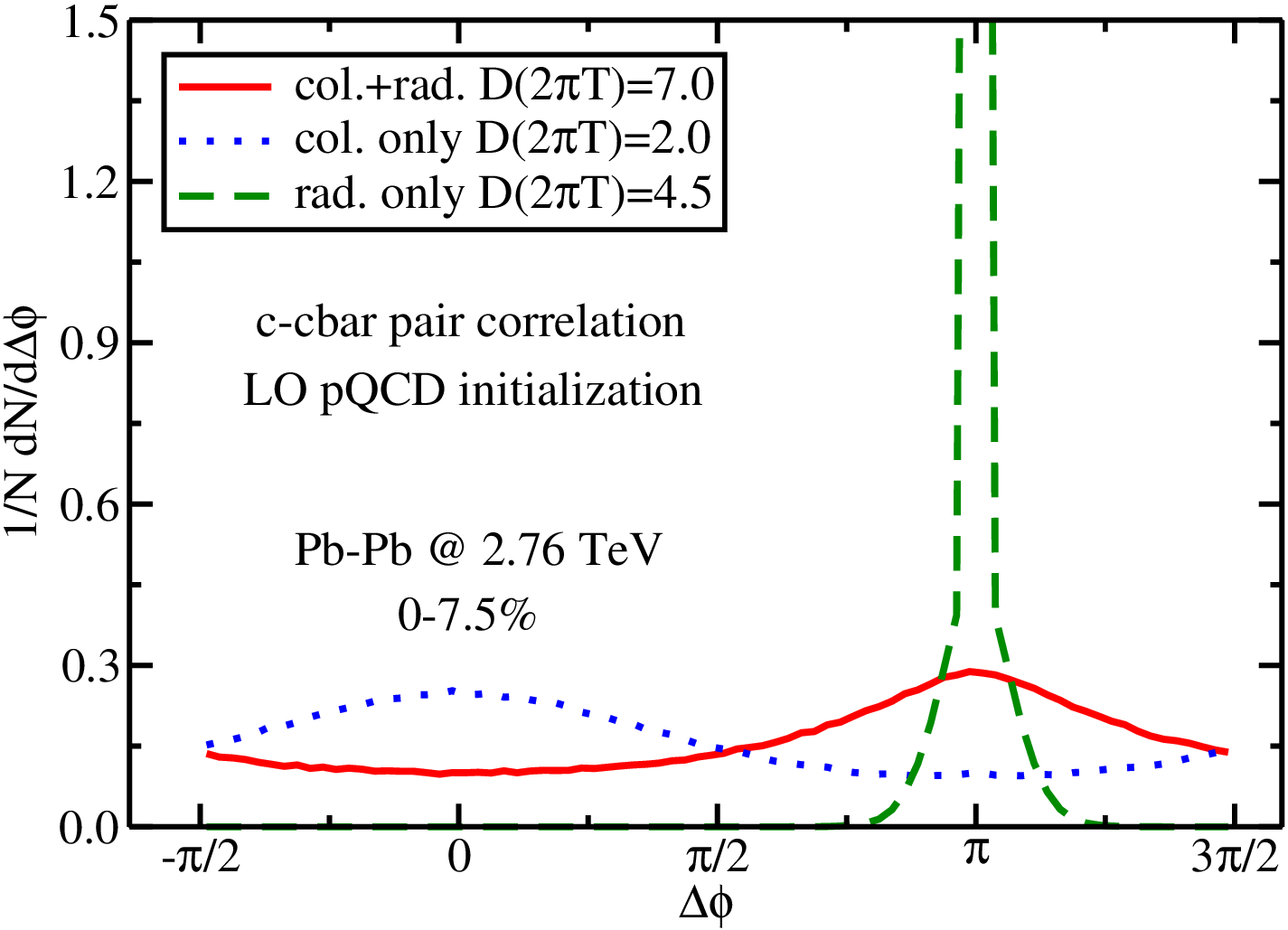}
\caption{\label{fig:corr_compare}(Color online) Comparison of the angular correlation functions of $c\bar{c}$ pairs between different energy loss mechanisms.}
\end{minipage}
\end{figure}

\begin{figure}[tb]
\begin{minipage}{0.47\textwidth}
\includegraphics[width=0.96\textwidth,clip=]{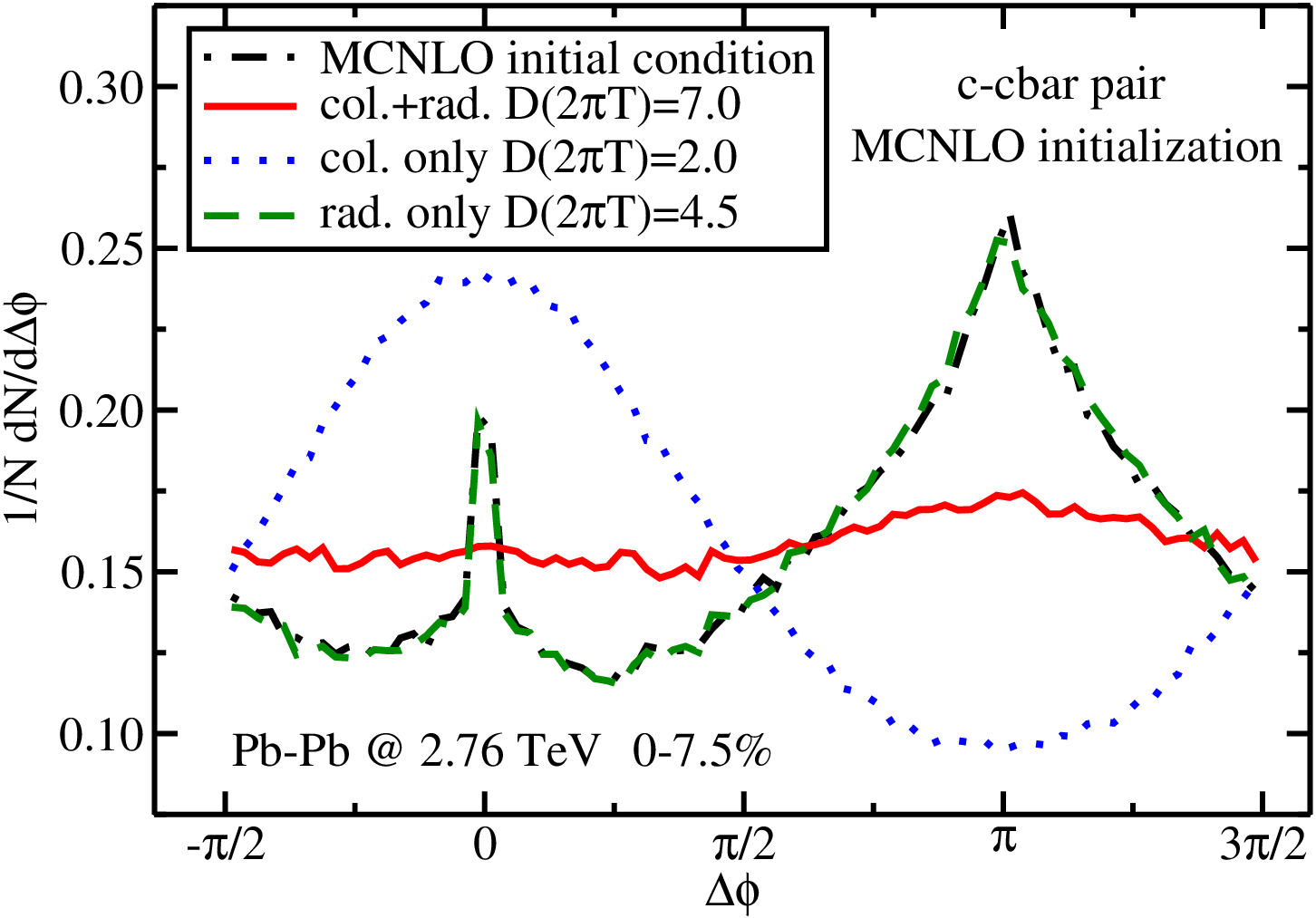}
\caption{\label{fig:ccbar-show}(Color online) Angular correlation functions of $c\bar{c}$ pairs with MCNLO+Herwig initial conditions.}
\end{minipage}\hspace{0.06\textwidth}
\begin{minipage}{0.47\textwidth}
\includegraphics[width=0.96\textwidth,clip=]{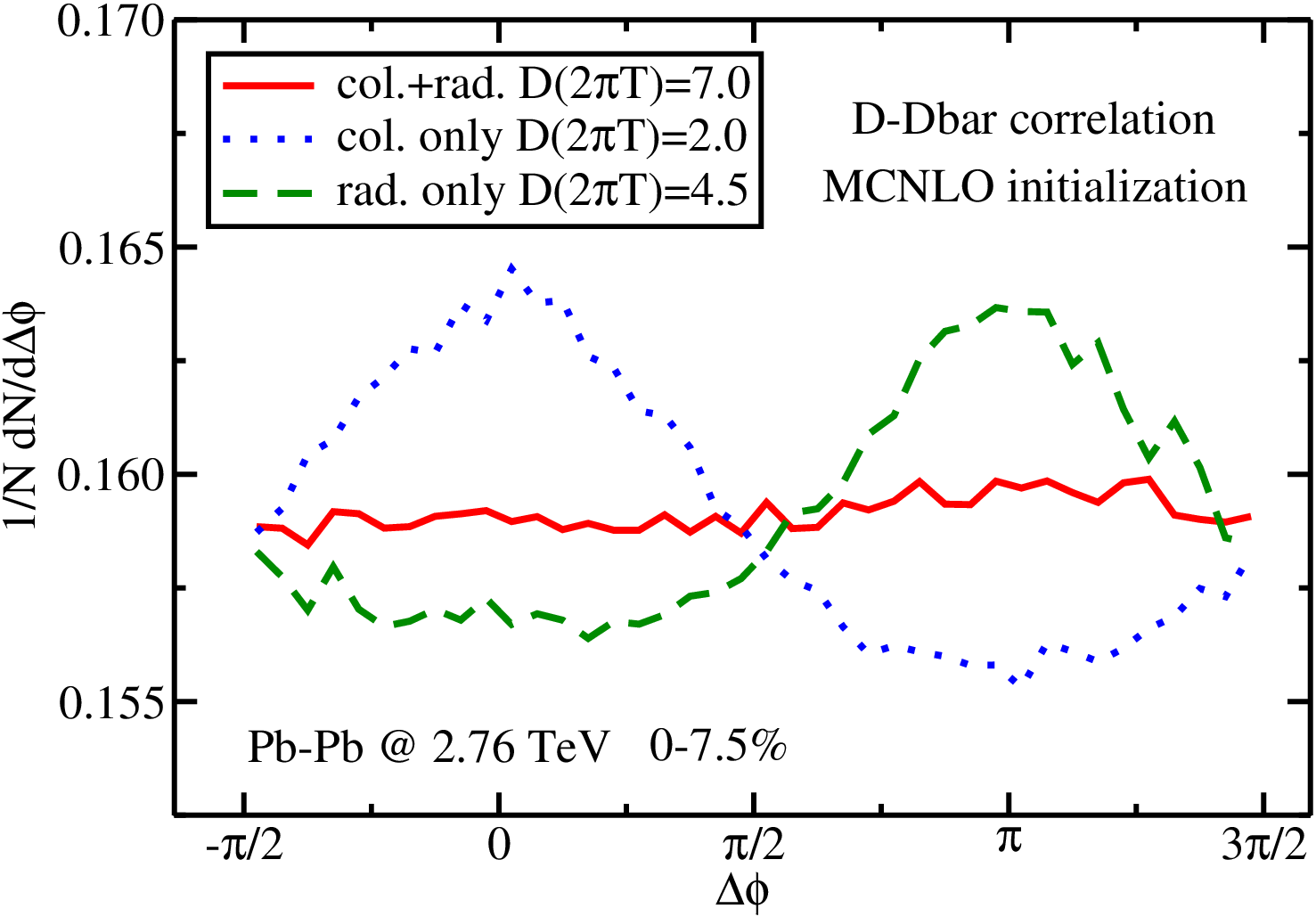}
\caption{\label{fig:DDbar-show}(Color online) Angular correlation functions of $D$-$\bar{D}$ with MCNLO+Herwig initial conditions.}
\end{minipage}
\end{figure}

We have shown good descriptions of $D$ meson suppression by including both collisional and radiative energy losses. However, in Fig.\ref{fig:RAA_fit_with_each_mechanism}, one sees that by tuning the transport coefficient, collisional or radiative energy loss alone also provide reasonable description of the data within the current experimental uncertainties. This implies the insufficiency of single particle spectra and motivate us to explore exclusive spectra of heavy flavor \cite{Nahrgang:2013saa,Cao:2013wka,Cao:2014pka}.

In Fig.\ref{fig:corr_compare}, we investigate the angular correlation functions of $c\bar{c}$ pairs after they traverse a QGP medium. We start with the leading order approximation of initial $c$ and $\bar{c}$ pairs and find that after they travel through the medium, the correlation function still peaks around $\pi$ if only gluon radiation is considered. However, a peak near 0 is observed if one only includes the collisional energy loss. This indicates that unlike the back-to-back initial state, low energy $c\bar{c}$ pairs tend to move collinearly in the end because of the boost by the radial flow of the medium.

In Fig.\ref{fig:ccbar-show}, we use an improved initialization for $c\bar{c}$ pairs -- the Monte-Carto next-to-leading-order production plus Herwig vacuum radiation \cite{Nahrgang:2013saa}. Similar to Fig.\ref{fig:corr_compare}, pure gluon radiation does not affect the angular correlation function, but pure scattering leads to a peak around 0. A more realistic analysis is implemented in Fig.\ref{fig:DDbar-show} where each $D$ meson is looped over all $\bar{D}$ within an event. We observe that the shapes of $D-\bar{D}$ correlation functions resemble those of $c\bar{c}$ pairs except for the existence of a large background of uncorrelated $D$ and $\bar{D}$. One may also find $D$-hadron correlation functions in our presentation though they may not as clean as those directly between heavy flavor pairs. Although these functions are dependent on different models, they may provide a deeper insight into heavy quark energy loss mechanisms if comparisons can be made between theory and future experiments.

\section{Conclusions}
We have developed a framework to describe the full-time evolution of heavy flavor in heavy-ion collisions. Within our framework, we have provided the $D$ meson suppression and flow consistent with most existing data, and explored the heavy-flavor-tagged angular correlation functions and found them useful in distinguishing between different energy loss mechanisms of heavy quarks inside the QGP. 

We thank the cooperations within the JET Collaboration. This work was supported by the U.S. Department of Energy Grant No. DE-FG02-05ER41367  and Natural Science Foundation of China (NSFC) under grant No. 11375072.





\bibliographystyle{elsarticle-num}
\bibliography{SCrefs}



\end{document}